\documentclass[a4paper]{VisionStyle}
\usepackage{epsfig}

\begin{document}

\outer\def\ltae {$\buildrel {\lower3pt\hbox{$<$}} \over
{\lower2pt\hbox{$\sim$}} $}                                  
\outer\def\gtae {$\buildrel {\lower3pt\hbox{$>$}} \over
{\lower2pt\hbox{$\sim$}} $}

\title{{\sl XMM-Newton EPIC observations of Her X-1}}

\author{G. Ramsay\inst{1} \and S. Zane\inst{1} \and 
  M. A. Jimenez-Garate\inst{2} \and J. W. den Herder\inst{3} \and
  C. J. Hailey\inst{4}} 

\institute{Mullard Space Science Lab, University College London,
Holmbury St. Mary, Dorking, Surrey, RH5 6NT, UK
\and 
MIT Center for Space Research, 77 Massachusetts Avenue, Cambridge,
MA 02139, USA
\and
Space Research Organisation of The Netherlands, Sorbonnelaan 2,
   3584 CA Utrecht, The Netherlands
\and
Columbia Astrophysics Laboratory, Columbia University, New York,
   NY 10027, USA
}

\maketitle 

\begin{abstract}

\noindent We present spin-resolved X-ray data of the neutron star
binary Her X-1. We find evidence that the Iron line at 6.4 keV
originates from the same location as the blackbody X-ray
component. The line width and energy varies over both the spin period
and the 35 day precession period. We also find that the correlation
between the soft and hard X-ray light curves varies over the 35 day
period.

\end{abstract}

\section{Introduction}

Her X-1 is a stellar binary system consisting of a neutron star and a
A/F secondary star. It has been extensively studied at many
wavelengths. Its main temporal observational characteristics are: the
spin period of the neutron star is $\sim$1.24 sec, the binary orbital
period is 1.7 day and there is a 35 day period seen in X-rays which has
been interpreted as due to a warped accretion disc precessing around
the neutron star. We have obtained three observations of Her X-1 made
using {\sl XMM-Newton} at 3 different epochs. In this paper we present
an analysis of spin resolved data obtained using the EPIC detectors.

\section{Observations}

\begin{figure}
\begin{center}
\setlength{\unitlength}{1cm}
\begin{picture}(8,5)
\put(-0.5,-1.4){\includegraphics{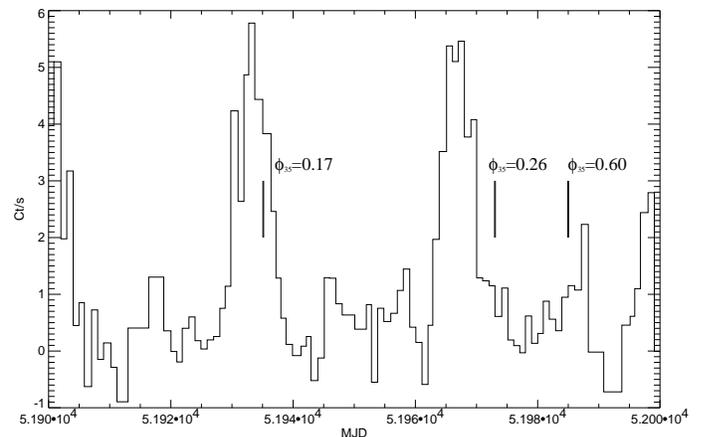}}
\end{picture}
\end{center}
\caption{The {\sl RXTE/ASM} (2-10)~keV light curve of
Her~X-1: the thick lines indicate the {\sl XMM-Newton} observations.}
\label{gramsay-c1_fig1} 
\end{figure}

The details of our observations are summarized in
Table~\ref{obslog}. To determine how these epochs relate to the 35~d
precession period ($\Phi_{35}$), we extracted the {\sl RXTE ASM}
(2-10)~keV quick-look light curve (Figure~\ref{gramsay-c1_fig1}). This
shows that the first observation took place close to maximum X-ray
brightness (`main-on'), while the third was close to the secondary
maximum (`short-on'). The second observation took place after the end
of the main-on state (`faint-state').

The EPIC detectors were configured in timing mode (apart from MOS2
which was in full frame mode and heavily piled-up). In the second
observation, data at the start and end of the observation were
excluded because of a higher particle background. Because of
difficulties in extracting background light curves and spectra in
timing mode, our data are not background subtracted. Since Her X-1 is
much brighter than the background in each epoch, we do not expect that
this will have a significant effect on our results. We have analysed
both EPIC MOS1 and EPIC pn data. The results from each detector are
consistent and here we show the results based on the pn data. Data
were processed using version 5.2 of the {\sl XMM-Newton} Science
Analysis System.

\begin{table}
\begin{tabular}{llllr}
\hline
Observation&Effective&EPIC PN&Orbital&35~d\\
Date & Exposure&Mean ct&Phase&Phase\\
     & (ksec) & (ct/s)& & ($\Phi_{35}$)\\
\hline
2001 Jan 26& 10& 398.4 &0.20--0.26&0.17\\
2001 Mar 4& 11& 21.4 &0.47--0.56&0.26\\
2001 Mar 17& 11& 54.3 & 0.52--0.60&0.60\\
\hline
\end{tabular}
\caption{Summary of the {\sl XMM-Newton} observations of Her X-1. For
the orbital phasing we use the ephemeris of Still et al (2001); for
the 35~day phasing we define phase 0.0 as point of the main high state turn-on.}
\label{obslog}
\end{table}

\section{Spin resolved light curves}

A barycentric correction and a correction for the motion of the
neutron star around the binary center of mass (using the ephemeris of
\cite{gramsay-c1:still01}) was applied to each photon. We then used a
discrete Fourier transform to obtain the spin period in each of the 3
epochs. Because of the problems with the clock on-board {\sl
XMM-Newton} these periods differ to some extent from that expected
from the recent spin history of this object (eg
\cite{gramsay-c1:oosterbroek01}). Since this study is concerned with
the relative phasing of the different energies rather than the precise
spin period, we are confident that since we have used the most
prominent peak in the power spectra our results are robust.

\begin{figure*}
\begin{center}
\setlength{\unitlength}{1cm}
\begin{picture}(6,12)
\put(-2.5,-1.5){\includegraphics{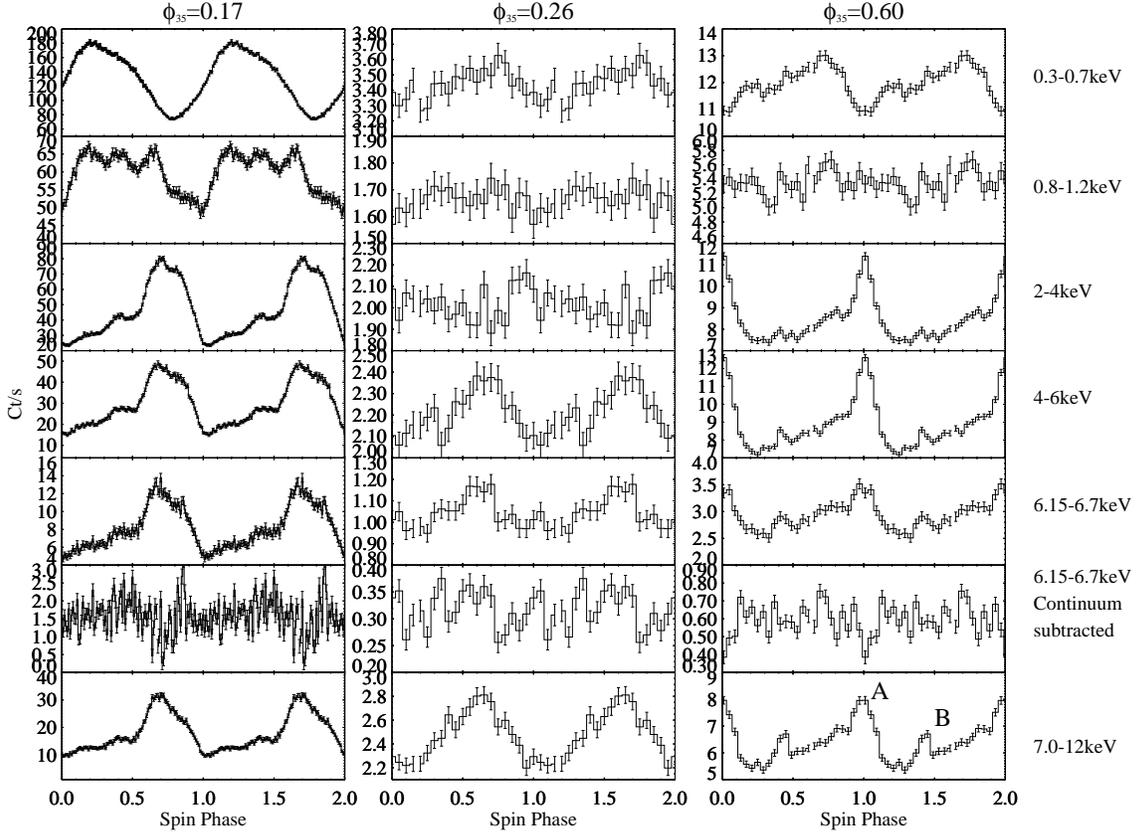}}
\end{picture}
\end{center}
\caption{The spin profiles for various energy bands for the three
epochs (left to right: the main-on, faint and short-on states. Because
of the uncertainty in the absolute timing, the spin phases are not on
the same absolute scale.}
\label{gramsay_c1-fig2} 
\end{figure*}

We show in Figure \ref{gramsay_c1-fig2} the energy-resolved
spin-folded light curves in our 3 observations. In the main-on state
($\Phi_{35}$=0.17) the softest (0.3--0.7 keV) and the harder ($>$2 keV)
light curves are anti-phased with some evidence that the minimum of
the continuum subtracted 6.15--6.7 keV light curve matches the peak of
the harder curves. In the faint-state ($\Phi_{35}$=0.26) the
modulation is much reduced, being greatest above 7 keV. The soft and
hard curves are approximately in phase. By the short-on state
($\Phi_{35}$=0.60) we find that the peak of the hard light curves
corresponds to the minimum of the softest.

\begin{figure}
\begin{center}
\setlength{\unitlength}{1cm}
\begin{picture}(8,5)
\put(-0.5,-0.5){\includegraphics{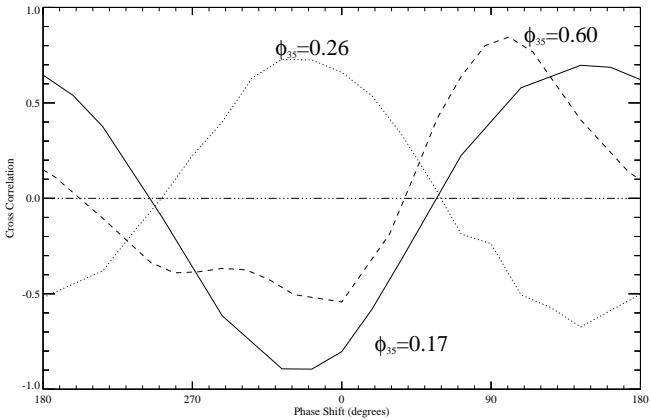}}
\end{picture}
\end{center}
\caption{The cross correlations for the soft and hard energy resolved
light curves at the 3 epochs.}
\label{gramsay_c1-fig3} 
\end{figure}

To investigate this more fully, we show the cross correlation of the
soft and hard light curves in Figure \ref{gramsay_c1-fig3}.  In the
main-on state we find that the most prominent cross correlation peak
shows that the soft and the hard light curves are strongly
anti-correlated, with only a small phase lag ($\sim-30^\circ$ or
330$^\circ$). We also find another (positive) peak at
($\sim130-170^\circ$) which corresponds approximately to the
separation between the peaks in the soft and hard light curves. The
faint state shows a positive correlation at small phase lag
($\sim330^\circ-340^\circ$) and a negative correlation at
$\sim130-160^\circ$. During the short-on state, we find the highest
correlation near $\sim100^\circ$. It is clear that the relative phase
shift between the soft and hard X-ray light curves varies as a
function of the 35~d precession period.

\section{Pulse phase spectroscopy}

\begin{figure*}
\begin{center}
\setlength{\unitlength}{1cm}
\begin{picture}(14,10)
\put(-2.5,-7.5){\includegraphics{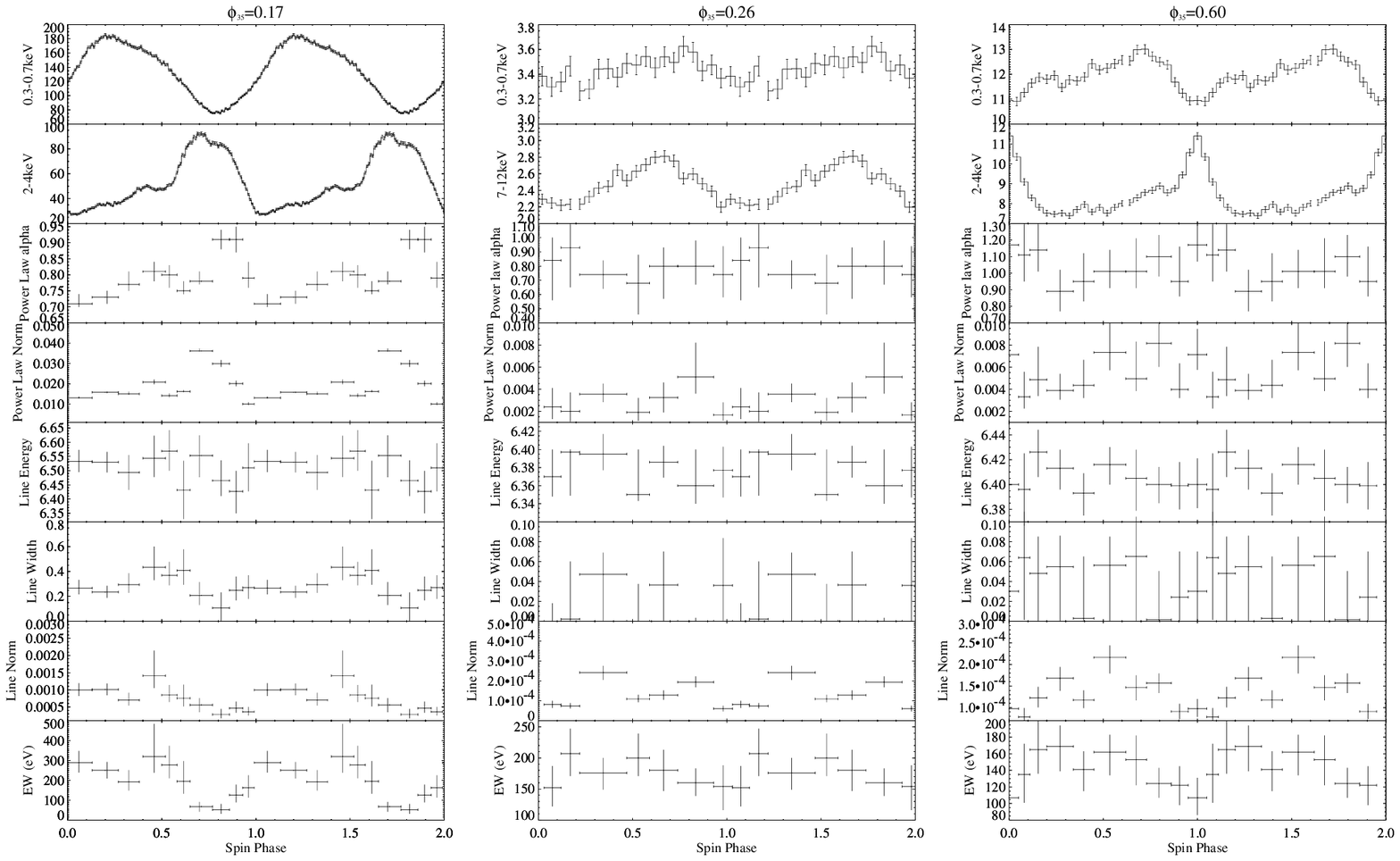}}
\end{picture}
\end{center}
\caption{The variation of the Fe line parameters as a function of the
spin phase, for the three different epochs. We also show the
normalization of the power law component and the intensity curves in
the soft (0.3-0.7~keV) and hard energy band.  The latter is taken to
be (2-4)~keV for $\Phi_{35}$=0.17, $\Phi_{35}$=0.60 and (7-12)~keV for
$\Phi_{35}$=0.26. The line width is the FWHM in keV.}
\label{gramsay_c1-fig4}
\end{figure*}

To search for a variation in the profile of the Fe line at 6.4~keV
over the spin cycle, we extracted spin resolved spectra. We ignored
energies less than 5 keV and more than 8 keV and used a Gaussian plus
power law model to model the spectra. We also determined the variation
in the power law normalisation by fitting an absorbed power law over
the range (2.0-6.1)~keV and (6.7-10)~keV, after fixing the slope to be
that of the best fit of the appropriate integrated spectrum. Results
are shown in Figure \ref{gramsay_c1-fig4}, together with the soft and
hard light curves.

At $\Phi_{35}$=0.26 there is no significant variation in the line
parameters. At phase $\Phi_{35}$=0.60 there is marginal evidence that
the minimum of the equivalent width corresponds to the maximum of the
intensity in the (2-4)~keV band. At $\Phi_{35}$=0.17 there is some
evidence that the line width reaches a minimum at the maximum of the
intensity of the (2-4)~keV light curve, but the most interesting
result is that the the equivalent width curve clearly follow the soft
X-ray light curve. This supports the idea that the 6.4~keV
fluorescence Fe line and the black body component originate from a
common region.

\section{Discussion}

\subsection{Energy resolved light curves}

Many features of the light curves are naturally explained within the
scenario proposed by \cite{gramsay-c1:scott00}. This model is based on the
obscuration of a multi-component X-ray beam by a counter-precessing,
tilted, twisted disk.  For simplicity, the X-ray beam is assumed to be
decoupled from the disk and is axisymmetric. One of the main features
of this model is that it ascribes the variations observed in the pulse
profile over the 35~day cycle to occultation from the {\it inner} part
of the disk, whereas most of the previous investigations have assumed
an occultation from the {\it outer} boundary.

The overall situation is summarized in the bottom panel of Figure 8 in
\cite{gramsay-c1:scott00}, while their Figures 10 \& 11 illustrate the
evolution of the pulse profiles predicted during the main-on and
short-on respectively. There is a similarity between the multi-peaked
hard light curve at $\Phi_{35}=0.17$ and the model during the
progressive occultation of leading and trailing peaks of the hard
beam. At $\Phi_{35} \sim 0.27$, when the main components are occulted,
we only observe the survival of a broad, underlying modulation that is
attributed to the magneto-spheric emission. Since this component is
emitted from a larger region at some distance from the neutron star,
it is naturally expected to have a lower modulation as well as a broad
maximum.

The pulse profile close to the short-on is also similar to that
presented by \cite{gramsay-c1:scott00} at $\Phi_{35}\sim 0.58$ (see
their Figure~11).  In the EPIC data, we can in fact recognize a main
peak ``A'' as well a small peak ``B''
(Figure~\ref{gramsay_c1-fig2}). However, because the notch ``B'' is
the hardest feature, spectral considerations suggest that this maximum
is associated with the small hard peak and the feature ``A'' with the
soft peak discussed by \cite{gramsay-c1:scott00}. If this is the case,
``B'' is actually due to direct emission from the pencil beam, while
``A'' is the radiation redirected into the fan beam from the antipodal
accretion column.

\subsection{The shift between the soft and hard curves}

Given the complexity of the source, pulse-phase spectroscopy is of
paramount importance to separate the different spectral components
observed in Her X-1. Using $Einstein$ and $BeppoSax$ data it has been
shown that, during the main-on state, the maximum of the thermal
component and the power law components are shifted by $\sim
250^\circ$. The situation is less consistent as far as the 6.4~keV Fe
K line is concerned: \cite{gramsay-c1:choi94} have shown that its
intensity is modulated in phase with the soft emission, suggesting a
common origin while \cite{gramsay-c1:oosterbroek00} have found it
correlated with the hard (power law) emission.

The shift in phase between the hard and soft emission can be explained
if the latter results from re-processing of hard X-rays in the inner
part of the accretion disk. If a non-tilted disk intercepts (and
re-processes) a substantial fraction of the hard beam from the neutron
star, the expected phase difference between direct and reflected
component is $180^\circ$.  Therefore, the value determined using
$Einstein$ and $BeppoSax$ data has been associated with the disk
having a tilt angle. If the tilt of the disk changes with the phase
along the 35~day cycle (as predicted by the precessing disk models,
see \cite{gramsay-c1:gerend76}) the shift in phase should therefore
vary with $\Phi_{35}$. However, both Einstein and Sax data were
obtained at the same $\Phi_{35}$.

The phase shift derived from {\sl XMM-Newton} data main-on state data
are considerably different from previous observations made in the
main-on state and continues to change during the other two
observations. This suggests that we are observing, a {\it substantial
and continuous variation in the tilt of the disk}, which is what we
would expect from a system which had a precessing accretion disc. It
should be noted that the interpretation of the phase shift observed at
the short-on may be affected by a systematic error, depending on
whether during the observation the soft peak ``A'' is higher than the
small hard peak ``B'' or vice-versa.

\subsection{The Fe 6.4 keV line variation}
\label{line}

At $\Phi_{35}=0.26$, there is little evidence for a significant
variation in the Fe line parameters, while at $\Phi_{35}=0.60$, there
is some evidence that the variation of the equivalent width of the Fe
line is in anti-phase with the (2--4)~keV intensity curve and follows
the general shape of the (0.3--0.7)~keV intensity curve.  At
$\Phi_{35}=0.17$, we find that the soft flux below 0.7~keV, the line
normalization, the line width and the equivalent width all exhibit a
common minimum at $0.7<\phi_{spin}<0.9$, which, in turn, is shifted
with respect to that of the hard emission. This supports the idea that
the 6.4~keV Fe line originates from fluorescence from the relatively
cold matter of the illuminated spot where the soft emission is
reprocessed.

We have also found evidence for a variation in the Fe line broadening
over the 35~day period. In addition, the energy of the Fe line suggests
that the Fe line emission originates from low ionisation species (Fe
XIV or less) in the low and short-on state observations, whereas in
the main-on the observed Fe K centroid energies ($6.52 \pm 0.03$~keV
for MOS and $6.50 \pm 0.02$~keV for PN) correspond to Fe XX-Fe XXI.

Taking this into account, the true centroids deviate by $\sim7 \sigma$
from the 6.40~keV neutral value. This suggests two possible
explanations for both the line broadening and the centroid
displacement: 1) an array of Fe K fluorescence lines exists for a
variety of charge states of Fe (anything from Fe I-Fe XIII to Fe
XXIII); 2) Comptonization from a hot corona with a significant optical
depth for a narrower range of charge states centered around Fe XX.

The Fe line broadening may also be explained in terms of Keplerian
motion, if the inner disk (or some inner region) comes into view
during the main-on state. If this is the case, at $\Phi_{35}=0.17$ the
Keplerian velocity will be $\sim 13000$ km/sec. This, in turn,
corresponds to a radial distance of $\sim 4 \times 10^8$~cm (for a
neutron star of 1.4\sun), which is close to the magneto-spheric radius
for a magnetic field of $\sim 10^{12}$~G.

\begin{acknowledgements}

This paper makes use of quick-look results provided by the ASM/RXTE
team whom we thank. We are grateful to Mariano Mendez for stimulating
comments and Vadim Burwitz for useful discussions.

\end{acknowledgements}

\end{document}